\documentclass[a4paper,12pt]{article}
\pdfoutput=1 

\usepackage{jcappub} 
\usepackage[T1]{fontenc}
\usepackage[latin1]{inputenc}
\usepackage{amsmath, amsfonts, amssymb, hyperref, graphicx, multirow, subfigure, bbold, bm}
\hypersetup{colorlinks=True,citecolor=blue}
\usepackage{soul}
\usepackage[usenames,dvipsnames,svgnames,table]{xcolor}
\usepackage{hyperref}
\usepackage{natbib}
\usepackage{array}
\usepackage{aas_macros}
\usepackage{epstopdf}
\pdfoutput=1

\graphicspath{ {"./figures/"} }

\title{\boldmath Measuring our velocity from fluctuations in number counts}

\author[a,1]{Nidhi Pant,\note{Corresponding author.}}
\author[b]{Aditya Rotti,}
\author[a]{Carlos A.P. Bengaly,}
\author[a,c]{Roy Maartens}

\affiliation[a]{Department of Physics \& Astronomy, University of the Western Cape, Cape Town 7535, South Africa}
\affiliation[b]{Jodrell Bank Centre for Astrophysics, University of Manchester, Manchester M13 9PL, United Kingdom}
\affiliation[c]{Institute of Cosmology \& Gravitation, University of Portsmouth, Portsmouth PO1 3FX, United Kingdom}

\abstract{
Our velocity relative to the {cosmic microwave background} (CMB) generates a dipole from the CMB monopole, which was accurately measured by COBE. {The} relative velocity also modulates and aberrates the CMB fluctuations, generating a {small}  signature of statistical isotropy violation in the covariance matrix. This signature was first measured by Planck 2013. Galaxy surveys are similarly affected by a Doppler boost. The dipole generated from the number count monopole has been extensively discussed, and measured (at very low accuracy) in the NVSS and TGSS radio continuum surveys. {For the first time,} we present an analysis of the Doppler imprint on the number count fluctuations, using {the} bipolar spherical harmonic formalism to quantify these effects. Next-generation wide-area surveys with a high redshift range are needed to detect the small Doppler signature in number count fluctuations. We show that radio continuum surveys with {the SKA should enable a detection at $\gtrsim 3 \sigma$ in Phase 2, with marginal detection possible in Phase 1.}}

\begin{document}
\maketitle
\flushbottom
\include{./todolist}
\section{Introduction}
\label{sec:intro}

Our motion relative to the cosmic microwave background (CMB) rest frame gives rise to a Doppler boost effect on the radiation. 
{The maximum anisotropy detected is at the level  $\sim 10^{-3}$, from the dipole.}
This large amplitude indicates that we are moving relative to the CMB rest frame -- the  {one} in which the dipole vanishes. Assuming that the dipole is entirely due to Doppler boosting of the CMB monopole, the velocity was measured from the COBE experiment as $v=369 \pm 0.9\,$km/s towards galactic coordinates $(l,b)=(263.99^{\circ}\pm 0.14^{\circ}, 48.26^{\circ}\pm 0.03^{\circ})$~\citep{Kogut1993,Hinshaw2009}.

A Doppler boost generates not only a dipole {anisotropy} from the CMB monopole, but also affects the primordial fluctuations of the CMB {by generating a distinctive statistical isotropy violation signature of aberration and modulation.}. The effects are small -- a  $\sim 10^{-3}$ correction to $\sim 10^{-5}$ fluctuations.
There are two contributions~\citep{Kosowsky2011,Amendola2011, Notari2012, Planck2014}: \begin{itemize} 
\item Doppler modulation  amplifies the fluctuations in the boost direction, and suppresses them in the {opposite} direction {(the same effect generates the kinematic dipole from the CMB monopole)} 
\item {Doppler aberration deflects}  the arrival direction of photons towards the boost direction, distorting the anisotropy pattern. 
\end{itemize} 

These imprints on the CMB fluctuations were not detectable with COBE and WMAP, but Planck's high angular resolution and sensitivity and low noise levels allowed the first measurement of this velocity signature in 2013~\citep{Planck2014} (confirmed in Planck 2015~\citep{2016A&A...594A..16P}).
Our motion also generates a dipole asymmetry in observed number of clusters which has been studied in \citep{Chluba2005}.
Our motion affects not only the CMB but also galaxy surveys. According to the standard model, the rest frames of the CMB and of matter should coincide. This provides a critical test of statistical isotropy, as first pointed out by~\citep{Ellis1984}.  

Radio continuum surveys have been used as an alternative independent probe to extract the cosmic kinematic dipole.
Pioneering tests were performed by~\cite{Baleisis1998}, but the detection of a kinematic dipole in NVSS radio number counts was only confirmed later by~\citep{Blake2002}. Subsequent analyses revealed a $>2\sigma$ discrepancy between the predicted and measured amplitudes of the velocity~\citep{Singal2011,Gibelyou2012,Rubart2013,Rubart2014,Tiwari2014,Tiwari2015,Tiwari2016,Colin2017,Bengaly2018}. The nonlinear effects of local large-scale structure and the allowance for flux density and calibration errors have not resolved the discrepancy~\citep{Rubart2014,Bengaly2018}.
Current radio surveys do not have sufficient number density, sensitivity and resolution for a robust determination of the dipole amplitude and direction from the Doppler effect on the monopole.
The SKA is forecast to measure this amplitude and direction with sufficient accuracy to detect significant deviations from theory~\citep{Schwarz2015,SKA}.

The measurement of the Doppler boost vector using the method described here is less susceptible to local structure bias.
The unprecedented angular resolution and sensitivity of the SKA continuum survey  will facilitate the detection of the much weaker Doppler effect on the number count fluctuations. This has not previously been investigated. Here we present for the first time a forecast for such a detection in future SKA radio continuum surveys.

\section{Doppler effects on number counts}
\label{sec:doppler}

The observer's motion relative to the rest frame of radio sources gives rise to Doppler boost effect on number count maps. {The largest effect is on the monopole, leading to the kinematic radio dipole, which has already been detected with higher amplitude in existing radio continuum surveys.} In this paper, we explore the effect of boosting on the fluctuations in the number counts. 

A photon with a frequency $\nu$ moving in direction $\bm{n}$ in the {CMB rest frame}, will be observed at  frequency $\nu^{\prime}$ and direction $\bm{n}^{\prime}$ by a {Solar System} observer moving with velocity $\bm{v}=\bm{\beta}c$~\citep{2018JCAP...01..013M}: 
\begin{eqnarray}
\nu^{\prime}&=&\gamma\big(1+ \bm{n} \cdot \bm{\beta}\big)\, \nu,\\
\bm{n}'&=&\bm{n}+\bm{\beta}-\big(\bm{n} \cdot \bm{\beta}\big)\bm{n}, \label{nn'1}
\end{eqnarray}
where we neglect terms of $O(\beta^2)$. Using the identity $\bm{\nabla}\big(\bm{n} \cdot \bm{\beta}\big)=\big(\bm{\beta}\cdot\bm{\nabla}\big)\bm{n}=\bm{\beta}-\big(\bm{n} \cdot \bm{\beta}\big)\bm{n}$ for the constant vector $\bm{\beta}$ and the unit radial vector $\bm{n}$, we can rewrite \eqref{nn'1} as
\begin{eqnarray}
\bm{n}'&=&\bm{n}+\bm{\nabla}\big(\bm{n} \cdot \bm{\beta}\big). \label{nn'}
\end{eqnarray}
The redshifts and solid angles observed in the two frames are related by
\begin{eqnarray}
1+z^{\prime}&=&(1+z)\big(1-\bm{n} \cdot \bm{\beta} \big),\\
d\Omega^{\prime}&=&\big(1-2\bm{n} \cdot \bm{\beta} \big)\,d\Omega\,.\label{dom}
\end{eqnarray}

For radio sources, the flux density is assumed to have a power-law dependence on frequency~\citep{Ellis1984}:
\begin{eqnarray}\label{eq:DPL}
S\propto \nu^{-\alpha},
\end{eqnarray}
where $\alpha$ is the spectral index of the source.  For the purposes of computing the effect of a Doppler boost, the observed number ${\cal N}$ of sources per solid angle above the flux density threshold is usually approximated by a power law:
\begin{equation}\label{eq:numbercnt-flux}
{\cal N}\equiv  \frac{dN}{d\Omega}(> S) \propto S^{-x }.
\end{equation}
For $\alpha$ and $x$, we follow the SKA1 Redbook~\citep{SKA}  and assume
\begin{equation}
{\alpha=0.75\,, \quad x=1\,.}
\end{equation}
A Doppler boost changes the observed number count as follows~\citep{Ellis1984}:
\begin{equation}
{\cal N}'(\bm{n}')={\cal N}(\bm{n})\big[1+A \,\bm{n}\cdot \bm{\beta}\big], \quad \quad A =2+x (1+\alpha)=3.75\,. \label{NN'}
\end{equation}
{In the {dipole modulation} amplitude $A$, the {factor of} 2 arises from the boost effect on solid angle, given by \eqref{dom}, while $x(1+\alpha)$ encodes the boost effect on flux density. $\alpha$ and $x$ depend on source populations, frequencies and flux limits, so the exact value of $A$ will be known once the survey is operational. For our forecasts we use SKA1 Redbook values.} 

We now separate the fluctuations in number counts from the background (or average) value $\bar{\cal N}$. Using \eqref{nn'} in \eqref{NN'}, we find that
\begin{equation}\label{eq:2a}
\delta {\cal N}' (\bm{n}')= \bar{\cal N}\,A\,{\bm{n}'}\cdot\bm{\beta}+\delta {\cal N}\big(\bm{n}'-\bm{\nabla}({\bm{n}'}\cdot\bm{\beta})\big)\big[1+{A \,\bm{n}'\cdot \bm{\beta} }\big].
\end{equation}
Note that ${\bm{n}'}\cdot\bm{\beta}={\bm{n}}\cdot\bm{\beta}$ to $O(\beta)$.
Defining the fractional observed number count contrast $\delta_{{\cal N}}=\delta {\cal N}/\bar{\cal N}$, {and neglecting terms of $O(\beta^2)$}, we can write \eqref{eq:2a} as
\begin{equation}\label{eq:2}
\delta_{{\cal N}'} (\bm{n}')=\delta_{\cal N}(\bm{n}')+ A\,{\bm{n}'}\cdot\bm{\beta}   + \delta_{\cal N}(\bm{n}')A\,{\bm{n}'}\cdot\bm{\beta}
-{\nabla}^i({\bm{n}'}\cdot\bm{\beta})\, \nabla_i \,\delta_{\cal N}(\bm{n}').
\end{equation}
The second term on the right of (\ref{eq:2}) is the Doppler modulation of the monopole. 
The third term gives the Doppler modulation of
 the number count fluctuations. The last term arises from Doppler aberration  
of the number count fluctuations.
 
Up to now, attempts to measure our velocity relative to the rest frame of radio sources have been limited to measuring the second term on the right of \eqref{eq:2}~\citep{Blake2002, Singal2011,Gibelyou2012,Rubart2013,Rubart2014,Tiwari2014,Tiwari2015,Tiwari2016,Colin2017,Bengaly2018}.  
The alternative is to extract our velocity via the last two terms on the right of (\ref{eq:2}), as has been done with Planck
~\citep{Planck2014,2016A&A...594A..16P} for the CMB. 
Assuming that the Universe is statistically isotropic in the rest frame, a Doppler boost generates off-diagonal correlations in the covariance matrix which leads to a breakdown of statistical isotropy in the observer frame.  
For the first time, we exploit this unique signature of the Doppler modulation and aberration of fluctuations in number count maps to study the feasibility of extracting the boost velocity and direction.

\section{Bipolar Spherical Harmonic formalism}
\label{sec:biposh}

A statistically isotropic Gaussian random field is completely characterised by the angular power spectrum $C_{\ell}$, or equivalently the diagonal of the harmonic space covariance matrix: $\langle a_{\ell m}\, a^*_{\ell' m'}\rangle = C_{\ell}\, \delta_{\ell \ell'}\, \delta_{m m'}$. The galaxy number count  map is rendered statistically anisotropic due to our motion relative to the rest  frame of radio sources, as discussed in \S\ref{sec:doppler}. Since this map has spatially varying statistical properties, it is not completely characterised by the angular power spectrum alone. In particular, information about the Doppler boost is encoded in the off-diagonal elements of the covariance matrix.

Without making any simplifying assumptions about the statistical isotropy of the {observed}  number count contrast, its two-point correlation function,
\begin{eqnarray}
{\xi'(\bm{n}_1' , \bm{n}_2')=\big\langle \delta_{{\cal N}'} (\bm{n}_1')\, \delta_{{\cal N}'}(\bm{n}_2') \big\rangle ,}
\end{eqnarray}
can be expanded in Bipolar Spherical Harmonic (BipoSH) functions \citep{2003ApJ...597L...5H}:
\begin{eqnarray}
{\xi'(\bm{n}_1' , \bm{n}_2')}= \sum_{\ell_1 \ell_2 LM}{\big(A^{LM}_{\ell_1 \ell_2}\big)'}\,\big\{\bm{Y}_{\ell_1}({\bm{n}_{1}'})\otimes \bm{Y}_{\ell_2}({\bm{n}_{2}'}) \big\}_{LM}\,.
\end{eqnarray}
Here $\big(A^{LM}_{\ell_1 \ell_2}\big)'$ are the coefficients of expansion and the BipoSH basis functions are an irreducible tensor product of spherical harmonics:
\begin{equation}
\big\{\bm{Y}_{\ell_1}(\bm{n}'_{1})\otimes \bm{Y}_{\ell_2}(\bm{n}'_{2}) \big\}_{LM}=\sum_{m_1 m_2}{\cal C}^{LM}_{\ell_1 m_1 \ell_2 m_2}\, Y_{\ell_1 m_1}(\bm{n}'_1)Y_{\ell_2 m_2}(\bm{n}'_2) \,,
\end{equation}
where ${\cal C}^{LM}_{\ell_1 m_1 \ell_2 m_2}$ are the Clebsch-Gordan coefficients. These basis functions form a complete orthonormal set on the space $S^2 \otimes S^2$. Owing to properties of the Clebsch-Gordon coefficients, the indices on the BipoSH coefficients satisfy the constraints
\begin{equation}
\left|\ell_{1} - \ell_{2}\right|\leq L \leq \ell_1 + \ell_2 \,, \quad -L \leq M \leq L\,. \label{eq:biposh_index_constraint}
\end{equation}
In this new language,  the BipoSH coefficients $A^{00}_{\ell_1 \ell_2}$ characterise the statistically isotropic component of the field and are related to the standard angular power spectrum via 
\begin{equation}
C_{\ell_1} =   \frac{(-1)^{\ell_1} }{ \sqrt{2 \ell_1+1}}\, A^{00}_{\ell_1 \ell_2}\,\delta_{\ell_1 \ell_2}  \,.
\end{equation}
The coefficients $A^{LM}_{\ell_1 \ell_2}$ for $L\neq0$ characterise the statistically non-isotropic component of the field. 

The BipoSH coefficients can be evaluated from the measured {$\delta_{\mathcal{N'}}(\bm{n}')$} by using the following estimator: 
\begin{equation}
{\big(\hat{A}^{LM}_{\ell_1 \ell_2}\big)'} = \sum_{m_1 m_2} {\cal C}^{LM}_{\ell_1 m_1 \ell_2 m_2}\, {a_{\ell_1 m_1}'a_{\ell_2 m_2 }'} \quad 
{\rm where}~~ {a_{\ell m}'}=\int d\Omega_{\bm{n}'}\, \delta_{\mathcal{N'}}(\bm{n}')\, Y_{\ell m}^*(\bm{n}')\,.
\label{eq:biposh_est}
\end{equation}
Properties of Clebsch-Gordan and spherical harmonic coefficients require that only modes satisfying $M=m_1 + m_2 $, $|m_1| \leq \ell_1 $, and $|m_2| \leq \ell_2 $, contribute to the sum in \eqref{eq:biposh_est}.

Note that the estimator is evaluated from the single measured realisation of the number count contrast map and we use the hat notation to denote the resultant BipoSH coefficients. The  coefficients corresponding to the correlation function $\xi'$ follow from an ensemble average: 
\begin{equation}
\big(A^{LM}_{\ell_1 \ell_2}\big)' = \big\langle \big( \hat{A}^{LM}_{\ell_1 \ell_2}\big)' \big\rangle\,.
\end{equation}

A model for a Doppler boosted number count contrast map is given by \eqref{eq:2}. Evaluating the two-point correlation function for this Doppler boost map, we can show that it generates only $L=1$ BipoSH coefficients (see~\citep{2016A&A...594A..16P,2015PhRvD..92h3015A} for details).  The BipoSH coefficients evaluated from the measured (boosted) map are
\begin{eqnarray}\label{eq:3}
\big(\hat{A}^{1M}_{\ell_1 \ell_2}\big)'=\hat{A}^{1M}_{\ell_1 \ell_2}+{\beta_{1M}}G^{1}_{\ell_1 \ell_2} + \mathcal{O}(\beta^2)\,,
\end{eqnarray}
{where $\hat{A}^{1M}_{\ell_1 \ell_2}$ are the rest-frame coefficients, from the first term in \eqref{eq:2}}, and 
\begin{equation}
\beta_{1M}=\int d\Omega_{\bm{n}}\,\big({\bm{n}}\cdot\bm{\beta}\big)\,{Y^{*}_{1M}}(\bm{n}) ,
\end{equation}
are the spherical harmonic coefficients of the Doppler boost field.
The shape function $G^{1}_{\ell_1 \ell_2}$ is completely characterised by the angular power spectrum $C_{\ell}$ of the isotropic component of the number count map~\citep{2016A&A...594A..16P,2015PhRvD..92h3015A}:
\begin{eqnarray}
G^{1}_{\ell_1 \ell_2}&=& \sqrt{\frac{(2\ell_1 +1)(2\ell_2 +1)}{12\pi}}\,{\cal C}^{10}_{\ell_1 0 \ell_2 0}\,\Big(A\,W^{1\,{\rm mod}}_{\ell_1 \ell_2}-W^{1\,{\rm abr}}_{\ell_1 \ell_2}\Big),\\
W^{1\,{\rm mod}}_{\ell_1 \ell_2}&=& C_{\ell_1}+C_{\ell_2}\,, \\ 
W^{1\,{\rm abr}}_{\ell_1 \ell_2}&=&C_{\ell_1}+C_{\ell_2}+ {1\over2}\big[\ell_1 (\ell_1 +1)-\ell_2 (\ell_2 +1)\big]\big(C_{\ell_1}-C_{\ell_2}\big).
\end{eqnarray}
Here $W^{1\,{\rm mod}}_{\ell_1 \ell_2}$ is the modulation contribution, from correlating the first and third terms in \eqref{eq:2}, and  $W^{1\,{\rm abr}}_{\ell_1 \ell_2}$ is the aberration contribution, from  correlating the first and fourth terms.
It is important to note that the spectral shape functions are determined only when the fiducial angular power spectrum  is well known.

The rest-frame coefficients vanish in an ensemble average, ${A}^{LM}_{\ell_1 \ell_2}=0$,  due to statistical isotropy. However, $\hat{A}^{LM}_{\ell_1 \ell_2}$  are non-vanishing and have the characteristics of noise,  which follows Gaussian statistics at multipoles $\ell \gtrsim 50$ \citep{2012PhRvD..85d3004J}. Note that this statement is only true for full-sky maps with equal measuring depths in all directions. Masking to avoid galactic contamination and varying observing depths in different portions of the sky can result in these rest-frame coefficients being significantly biased towards nonzero values. These biases need to be  modeled by carefully 
incorporating the effects in simulations of number count contrast maps.

Every pair of multipoles $\ell_1,\ell_2$ in \eqref{eq:3}  that satisfy the triangularity condition for $L=1$ provides an estimate of ${\beta}_{1M}$:
\begin{equation}
\hat{\beta}_{1M}={\big(\hat{A}^{1M}_{\ell_1 \ell_2}\big)'\over G^{1}_{\ell_1 \ell_2}} = \beta_{1M} + {\hat{A}^{1M}_{\ell_1 \ell_2}\over G^{1}_{\ell_1 \ell_2}}\,.
\end{equation}
An optimal estimator which maximises the signal to noise ratio is the sum of the inverse variance-weighted estimate of $\hat{\beta}_{1M}$ inferred from each distinct pair of multipoles $\ell_1, \ell_2$.
A minimum variance estimator for $\beta_{1M}$ is then given by~\citep{2015PhRvD..92h3015A}: 
\begin{eqnarray}
\hat{\beta}_{1M}=\Bigg[\sum_{\ell_1 \ell_2} \frac{G^{1}_{\ell_1 \ell_2}\big(\hat{A}^{1M}_{\ell_1 \ell_2}\big)'}{{\tilde C_{\ell_1}\tilde C_{\ell_2}}}\Bigg]\, N_1\,, 
\end{eqnarray}
{where the noise variance of the reconstructed Doppler field is}
\begin{eqnarray}
N_1=\Bigg[\sum_{\ell_1 \ell_2} \frac{(G^{1}_{\ell_1 \ell_2})^{2}}{{\tilde C_{\ell_1}\tilde C_{\ell_2}}}\Bigg]^{-1}\,, \quad \tilde C_{\ell} \equiv C_{\ell}+{\cal N }^{-1}\,. \label{recon-noise}
\end{eqnarray}
The inverse angular number density, ${\cal N }^{-1}$, is the Poisson noise, which depends on the flux limit of the experiment.
The noise variance $N_1$ is sourced primarily by  cosmic variance and Poisson noise and depends on the number of modes that contribute to the sum in \eqref{recon-noise}. 

The direction of the Doppler boost velocity is inferred from how the power is distributed among the $M$ modes. The amplitude {of the Doppler boost velocity} is given by the estimator,
\begin{equation}\label{b1m}
{\hat\beta^2}=\frac{9}{4 \pi} \left({1\over3}\,{\sum_M \big|\hat{\beta}_{1M}\big|^2}   - N_1 \right) \,.
\end{equation}
Here $N_1$ gives the bias on the Doppler power resulting from statistically isotropic skies, which needs to be subtracted to infer the true Doppler power. A statistically significant detection of the Doppler vector using this technique {requires}  the error on the bias {to be} significantly smaller than  $\beta^2$. 
{Assuming Gaussian\footnote{Strictly speaking the reconstruction noise follows $\chi^2$ statistics.}  statistics, the error on $N_1$ is}
\begin{equation} \label{recn}
\sigma_{N_1} = {1\over f_{\rm sky}}\, \sqrt{2\over3}\,N_1\,,  
\end{equation}
where $f_{\rm sky}$ is the fraction of sky covered.


\section{Radio continuum surveys with the SKA}
\label{sec:SKA_measurements}

Radio continuum surveys measure the integrated emission of radio sources in a single redshift bin for a particular frequency band, integrating the
source flux in the band, with rms noise
$S_{\rm rms}$. 
The survey with Phase~1 of
SKA is expected to have the following properties~\citep{SKA}:
\begin{eqnarray}
{{\rm SKA1:}\quad f_{\rm sky}\approx 0.47\,, \quad 350<\nu<1050\,{\rm MHz}\,,  
 \quad S_{\rm rms}\sim 1\,\mu{\rm Jy}\,,
 \quad 0< z\lesssim 5\,.}
\end{eqnarray}
{SKA2~specifications have not been formalised, but the rms noise is expected to be $\sim10$ times smaller and the sky coverage increased from 20,000 to 30,000\,deg$^2$:}
\begin{eqnarray}
{{\rm SKA2:}\quad f_{\rm sky}\approx 0.7\,,  \quad S_{\rm rms}\sim 0.1\,\mu{\rm Jy}\,.}
\end{eqnarray}
We assume the SKA2 frequency range is the same as SKA1.

The flux threshold should be $\sim10$ times the rms noise level, and we assume:
\begin{eqnarray}
{\rm SKA1:} && \quad{{\rm optimistic:} ~~  S>10\,\mu{\rm Jy}\,,\quad {\rm realistic:} ~~ S>20\,\mu {\rm Jy}\,,} \label{smin}\\
{{\rm SKA2:}} && { \qquad\qquad\qquad ~~ S>~1\,\mu{\rm Jy}\,.}
\end{eqnarray}
{SKA1 is expected to detect sources out to $z\sim 5$, and SKA2 to $z\sim 6$. We use a redshift range $0.02<z<4.98$ for our forecasts.}

\begin{figure}
\centering
\includegraphics[width=0.49\textwidth]{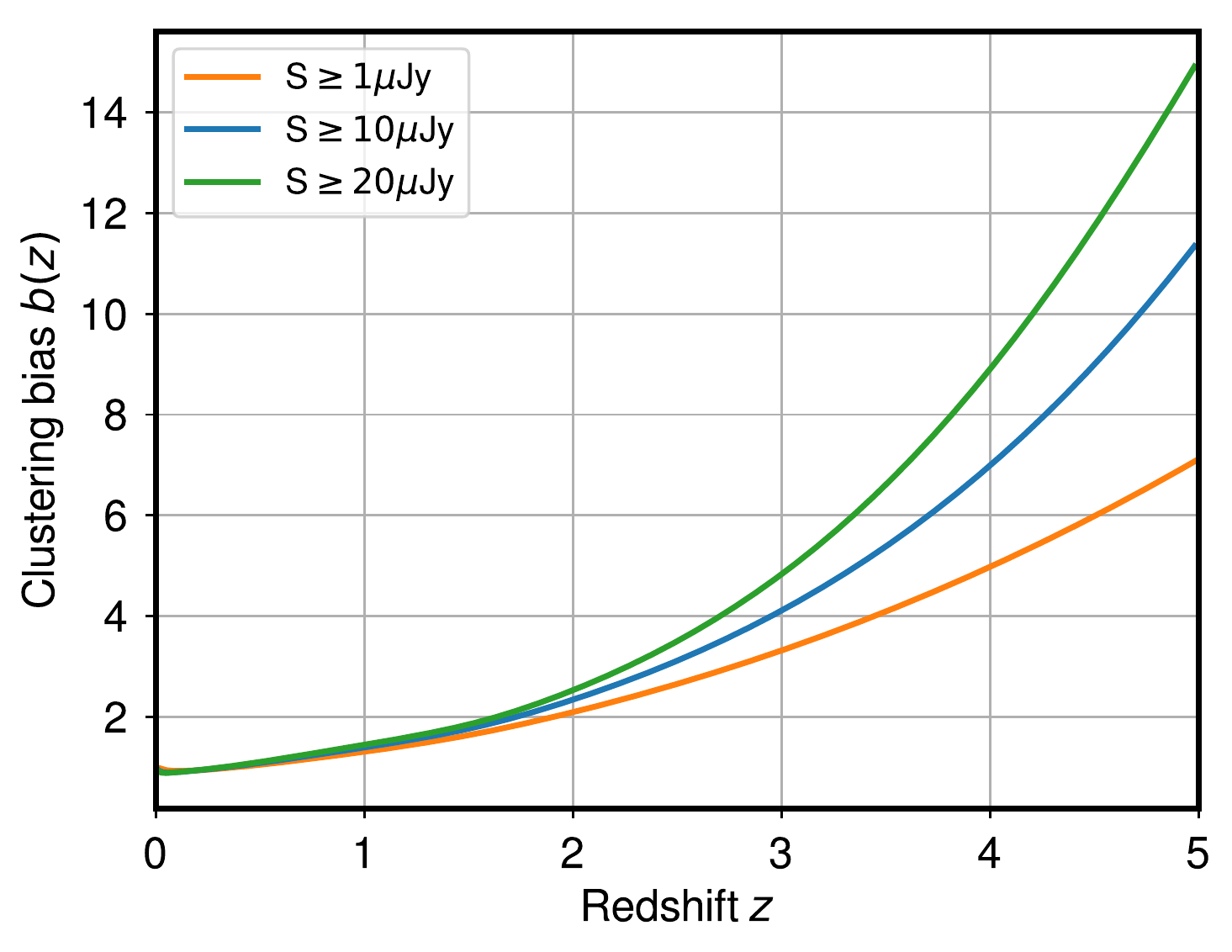}
\includegraphics[width=0.49\textwidth]{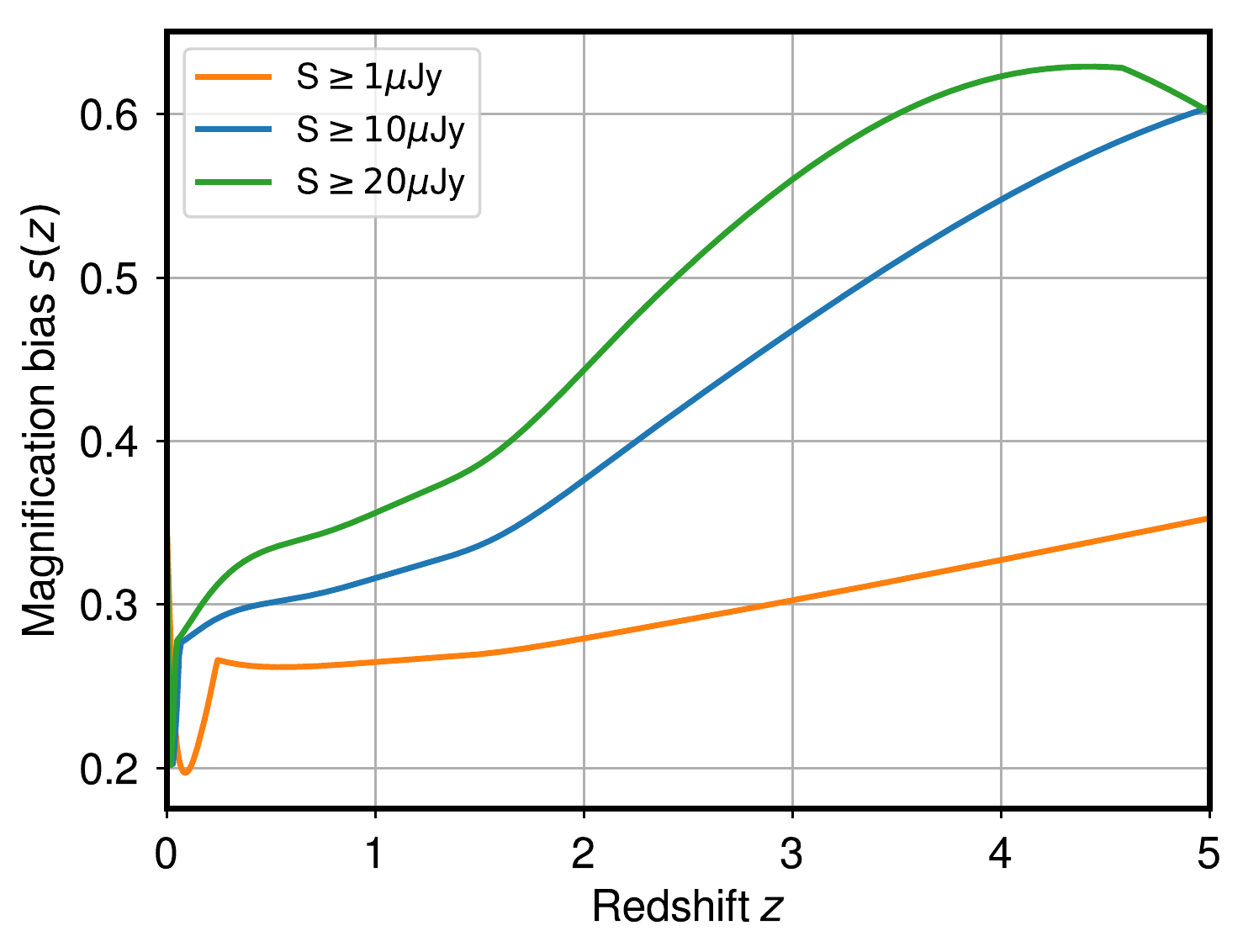}\\~\\
\includegraphics[width=.8\textwidth]{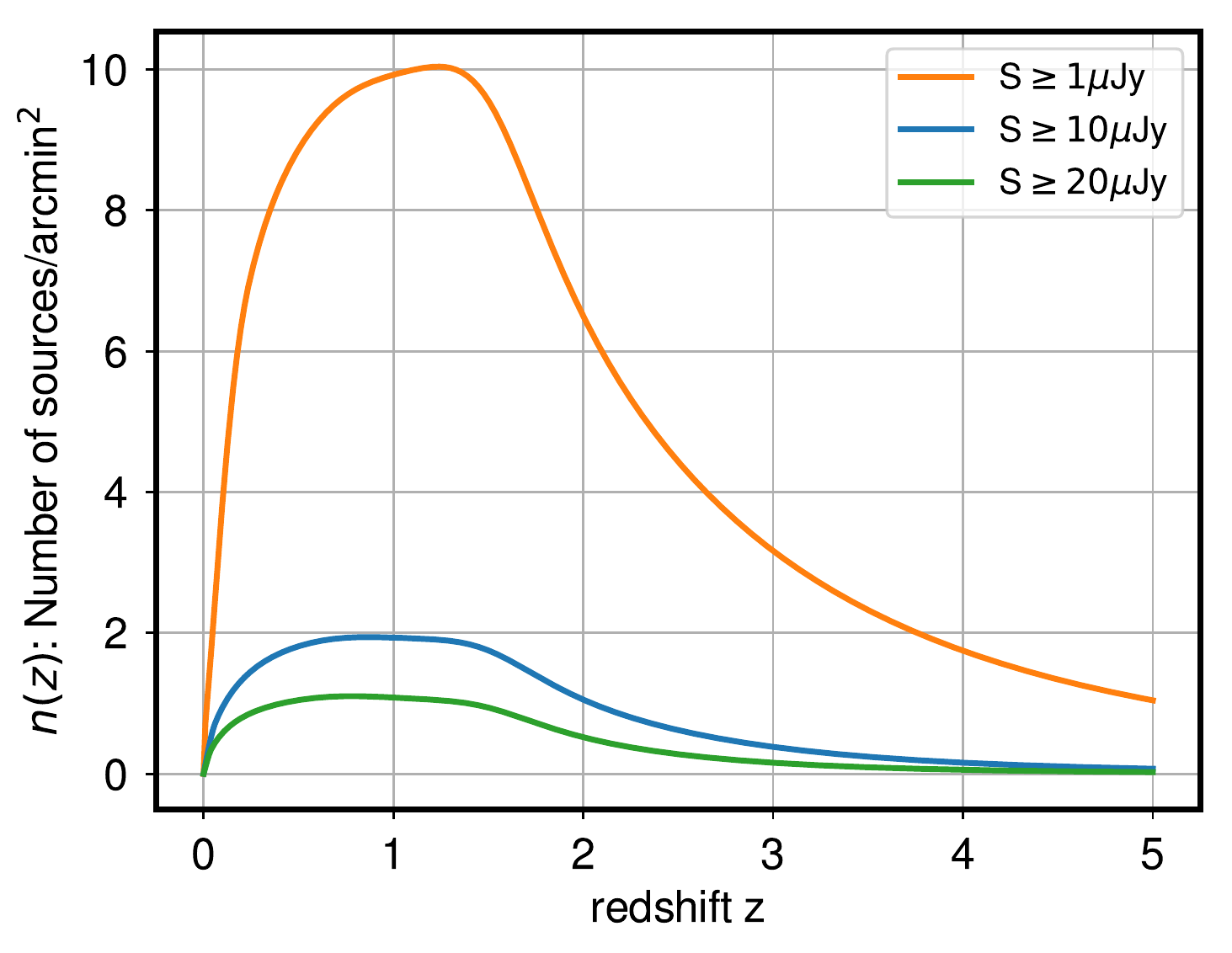}
\caption{{Clustering bias (top left) and magnification bias (top right) as a function of redshift. Number of sources per arcmin$^2$ per  redshift (bottom), for the flux thresholds 1, 10 and 20\,$\mu$Jy. }}
\label{fig:nz_bz_sz}
\end{figure}

An important quantity required for forecasts is the theoretical angular power spectrum of radio sources.
This is evaluated using {\sc CAMB Sources}~\citep{Challinor2011} which needs clustering  bias, $b(z)$, magnification bias, $s(z)$, and the redshift distribution of radio sources, $n(z)$ as input.
 The SKA Simulated Skies (S$^3$) database~\citep{Wilman2008} allows one to estimate the angular number densities at each redshift for the different types of radio sources and then estimate the total angular number density,  $n(z)$. 
Similarly, S$^3$ can be used to estimate the total clustering bias and magnification bias. However, in order to avoid having to query the full simulation, we follow the semi-analytic method of~\cite{Alonso2015}, which is based on~\citep{Wilman2008},  and we use their code to obtain $b(z)$, $s(z)$ and $n(z)$ for the continuum for different flux cuts.\footnote{http://intensitymapping.physics.ox.ac.uk/codes.html} {The number density and clustering bias used for the continuum to compute the fiducial angular power spectrum for galaxy number maps with different SKA configurations are depicted in Fig.~\ref{fig:nz_bz_sz}.}
\begin{figure}[!tbp]
\centering 
\includegraphics[width=0.49\textwidth]{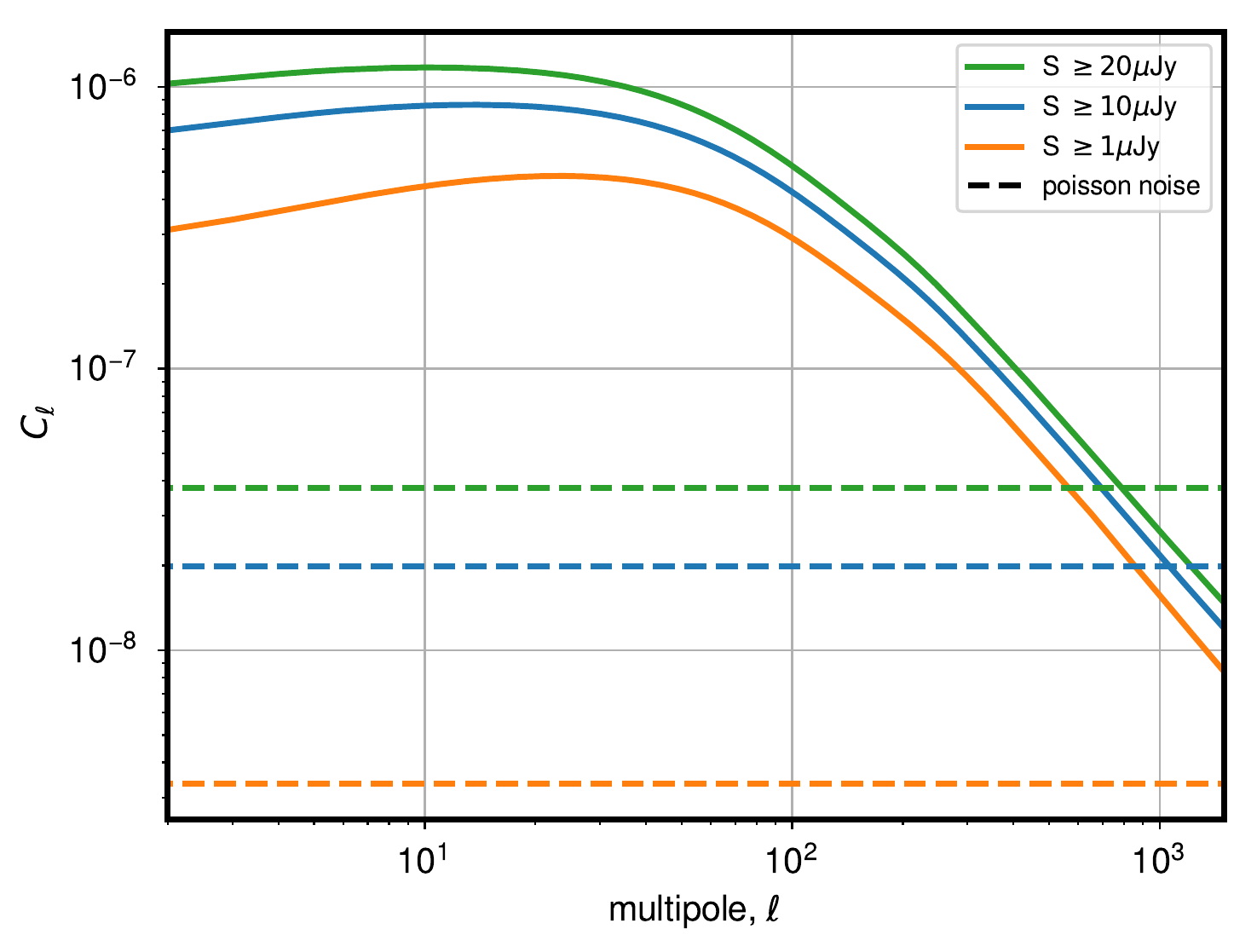}
\includegraphics[width=0.49\textwidth]{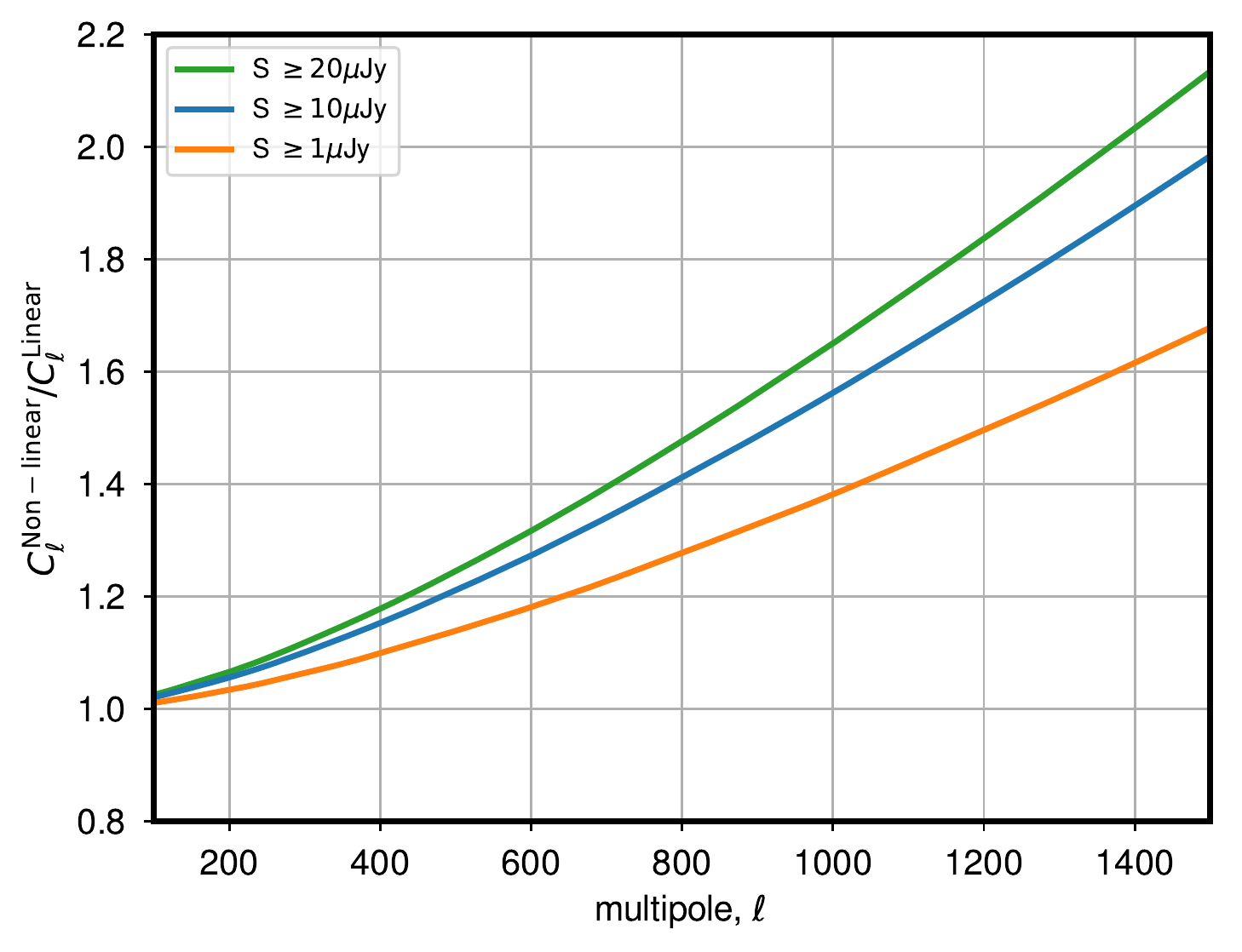}
\caption{\label{fig:spectrum} {\em Left:} Fiducial nonlinear  angular power spectra for SKA number counts maps   with their corresponding Poisson noise (dashed). {\em Right:} Ratio of nonlinear to linear power spectra.
}
\end{figure}

The reconstruction noise (\ref{recon-noise}) is obtained from a sum over multipoles and determines the $1\sigma$, $2\sigma$ or $3\sigma$ detection thresholds. The lower limit $\ell_{\rm min}$ of the sum  is determined by the sky area: SKA will allow  $\ell_{\rm min}=2$. The upper limit $\ell_{\rm max}$ is determined by the angular resolution, which is $\sim 0.5\,$arcsec in SKA1 and better in SKA2.

The angular power spectrum $C_\ell$ is computed with {\sc CAMB Sources}, using the updated Halofit~\citep{Takahashi2012} to compute nonlinear effects.   The fiducial cosmology is taken as the Planck 2015 best-fit flat $\Lambda$CDM model~\citep{Planck2016}. 
We integrate over the full redshift range $0.02<z<4.98$, with $n(z)$, $b(z)$ and $s(z)$ as shown in Fig.~\ref{fig:nz_bz_sz}, to obtain the theoretical projected 2-dimensional angular power spectrum.
However, in practice $\ell_{\rm max}$ will be much less than allowed by the resolution  -- since we need to  exclude nonlinear scales where theoretical predictions are unreliable.

{The resulting nonlinear angular power spectra and Poisson noise for SKA1 and 2 are shown in Fig.~\ref{fig:spectrum}.}
The different flux thresholds result in significant variations in the Poisson noise amplitude.
We also show the ratio of nonlinear to linear spectra in order to demonstrate the increase in power due to nonlinear effects.

\section{Forecasts for SKA}
\label{sec:forecasts}

For our forecasts, we have used SKA specifications, Planck 2015 best-fit flat $\Lambda$CDM model as fiducial cosmology along with astrophysical parameters $x$ and $\alpha$ from the SKA1 Red Book~\citep{SKA} to estimate errors on the reconstructed Doppler signal.
A statistically significant detection of the Doppler boost vector can be made if $\sigma_{N_1} \ll \beta^2$, as can be seen from (\ref{b1m}) and (\ref{recn}). The estimates are understandably very sensitive to the flux threshold as it determines the Poisson noise amplitude. Increased Poisson noise reduces the number of signal-dominated modes, and consequently raises the error on the reconstructed Doppler field.

{The high $\ell$ modes have a lower cosmic variance and hence contribute more to the signal, until Poisson noise takes over and the gains from adding more high multipoles diminish. The noise in the reconstructed Doppler field $\bm {n}\cdot \bm {\beta}$ therefore depends on how many modes with low cosmic variance there are, as shown in Fig.~\ref{fig:spectrum}. For this reason the reconstruction noise is lowered more by adding high $\ell$ modes which are signal-dominated. The low $\ell$ modes contribute relatively less to reducing the reconstruction noise, owing to larger cosmic variance. This reflects in the SNR estimates for the three flux threshold. The nonlinear increase in power in Fig.~\ref{fig:spectrum} occurs on scales where the Doppler signal is strong and shot noise is sub-dominant. Hence, including nonlinear effects reduces the maximum multipole required to reach SNR$\sim1$ or more. The gain is maximal for the lowest sensitivity case $S>20\,\mu$Jy,  since the nonlinear boost of power includes more signal-dominated modes for the estimation. In the highest sensitivity case $S>1\,\mu$Jy, this reduction is smallest since there are additional signal-dominated modes available due to a reduction in Poisson noise.
The largest scale $\ell_{\rm min}$ is important because the dipole $\bm{n}\cdot\bm{\beta}$ is a large-scale feature, and this is why the sky area is important.}
\begin{figure}[t!]
\centering 
\includegraphics[width=0.49\textwidth]{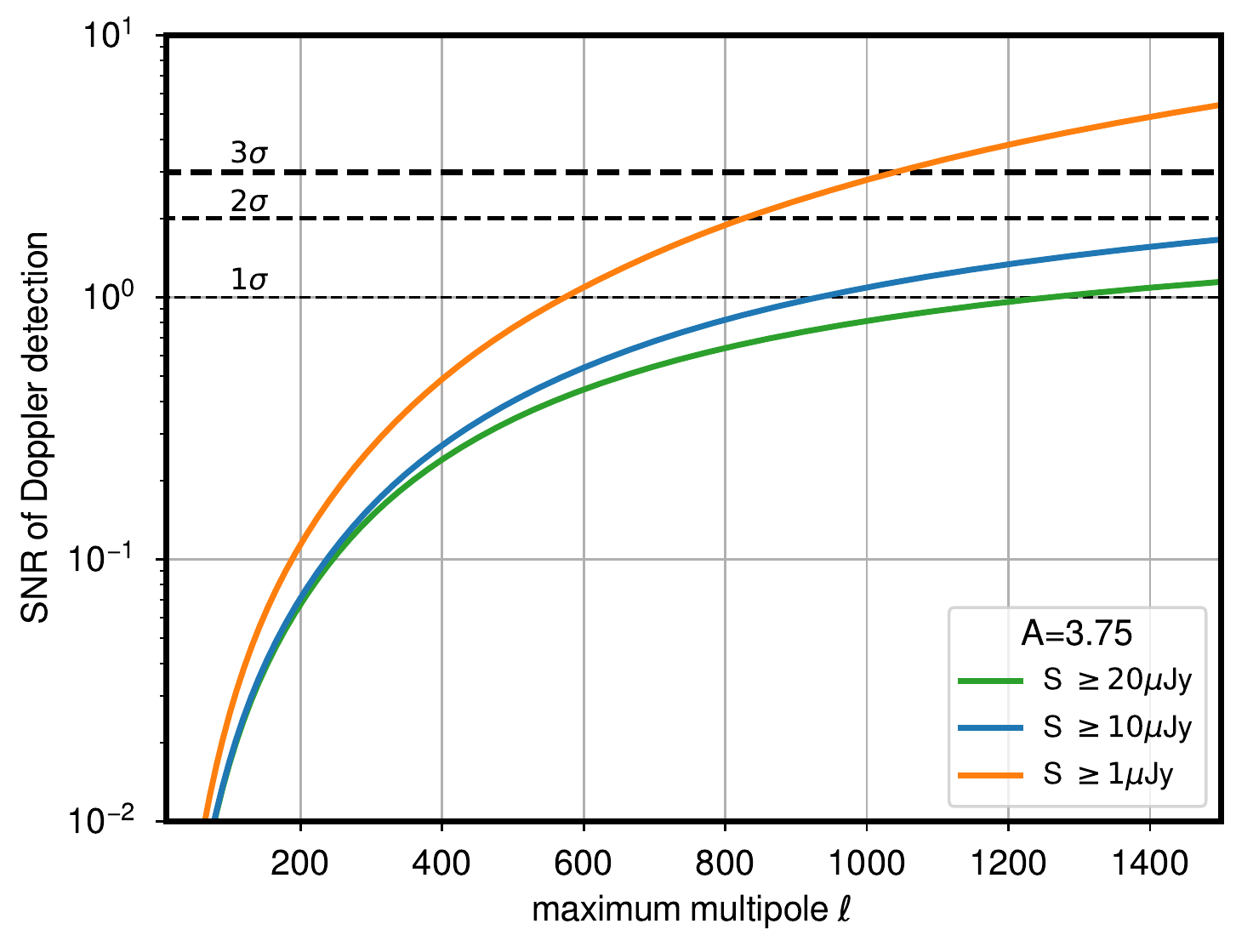}
\includegraphics[width=0.49\textwidth]{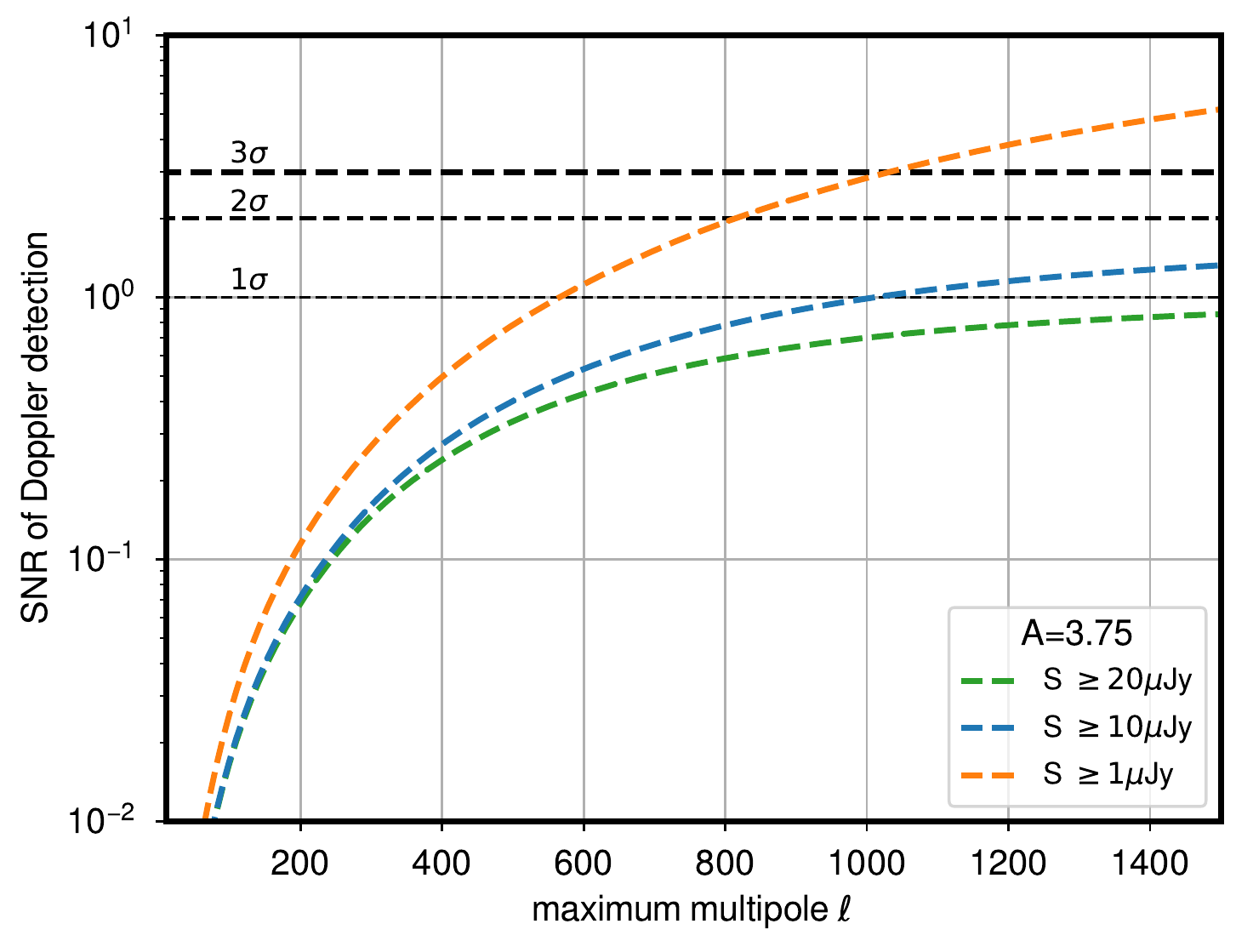}
\caption{ Signal to noise forecast to detect the Doppler boost from SKA number count fluctuations, using nonlinear (left) and linear (right) angular power spectrum.  }
\label{fig:10ujy}
\end{figure}

{In Fig.~\ref{fig:10ujy} we show the SKA signal to noise for detection of the Doppler signal at 1, 2 and 3$\sigma$ levels.}
{With SKA2,
the fiducial $A$ is predicted to deliver a $\gtrsim 3\sigma$ detection for $\ell_{\rm max} \gtrsim 1000$. SKA1 (optimistic) could make a $\gtrsim1\sigma$ detection for the fiducial $A$ with $\ell_{\rm max} \gtrsim 900$. In the realistic case, SKA1 does not make a detection unless $A$ is significantly higher than the fiducial value.}

{The forecasts depend on the value assumed for $A$, which itself depends on astrophysical properties of the source population. These properties will be determined when SKA observations are taken, but for forecasts, we need to make simplifying assumptions. We followed the standard assumption that $A$ is constant and we used the SKA1 Red Book value \eqref{NN'}.  
We do not model  uncertainties in the astrophysical parameters. A future study will be done to check the effect of such uncertainities on our estimates.}
Our forecasts also depend on nonlinear modelling of the fiducial power spectrum. We followed standard practice and used the updated 
Halofit~\citep{Takahashi2012} to estimate nonlinearities by emulating N-body codes. It is not  possible to perform a  rigorous accuracy test of such emulators, 
given that current N-body codes themselves disagree at the few percent level on nonlinear scales~~\citep{Schneider}. Updated Halofit agrees with the main N-body codes 
within $\sim5\%$ up to $k\sim10h\,$Mpc$^{-1}$~\citep{Schneider}. At $\ell\gtrsim1000$, the maximum $k$ that we rely on is at the minimum redshift $z=0.02$, and is given by 
$k=\ell r\gtrsim1000\, r(0.02) =17h\,$Mpc$^{-1}$ (where $r$ is comoving distance).  This is at the limit of Halofit accuracy, but there are also very few sources at such 
low $z$, with minimal contribution to the SNR.

Recently an improved emulator, the EuclidEmulator, has been developed and it agrees with the updated Halofit within a few percent on nonlinear 
scales~\citep{Knabenhans}. When data is taken in next-generation galaxy surveys (including SKA), 
accurate modelling of nonlinearities will require improved N-body codes and emulators,  as pointed out in~\citep{Schneider,Knabenhans}.

\section{Discussion and conclusions}
\label{sec:discussion}

{There is a potential `contamination' of the unboosted map by galaxy peculiar velocities and by the lensing magnification effect on number counts. For a continuum survey, the peculiar velocities (and therefore RSD) are nearly averaged out by projection of number counts onto the unit sphere giving rise to a few percent change in power spectra.}

Lensing magnification affects number counts through a change in solid angle and through magnification bias $s(z)$ (see Fig.~\ref{fig:nz_bz_sz}), which can move sources into or out of the map. This effect potentially leads to a bias on the estimate of the Doppler aberration effect, and in principle, the number count map should be de-lensed before applying the Doppler boost estimator. For SKA1, the lensing effect on continuum number counts is below the signal to noise, as shown in~\cite{Alonso2015} (see their Fig.~17), and can be safely neglected. For SKA2, the effect is above signal to noise unless the error on the magnification bias is $\gtrsim 20\%$. We computed the signal to noise forecast for detection of the Doppler boost, including both lensing and RSD effects.

In summary, fluctuations in number count maps are affected by our motion with respect to the CMB rest frame. For the first time,  we have proposed a method to extract the Doppler boost, {using the statistically anisotropic signature induced by modulation and aberration of fluctuations in number counts} from radio continuum surveys. Our forecasts indicate that SKA2 will be able to {detect} the Doppler field at a level that matches Planck at $\ell \sim 1000$, and improves on Planck if higher multipoles are accessed:
\begin{eqnarray}
\mbox{SKA2 detection:} \quad \gtrsim 3\sigma \quad \mbox{for}~~\ell_{\rm max} \gtrsim 1000\,.
\end{eqnarray}
For SKA in Phase 1, a marginal detection may be possible, as shown in Fig.~\ref{fig:10ujy}. 

Such a detection would provide an important further test of the Cosmological Principle, which dictates consistency  between the  CMB and the matter distribution.

\newpage
\subsection*{Acknowledgments}
We thank Debabrata Adak for early exploration in this work, and Mario Ballardini and Atsushi Taruya for useful discussions and inputs to this work. NP is supported by a Claude Leon Foundation Postdoctoral Fellowship.
NP, CB and RM acknowledge support from the South African SKA Project and the National Research Foundation of South Africa (Grant No. 75415). RM was also supported by the UK Science \& Technology Facilities Council (Grant No. ST/N000668/1).
We acknowledge the support of the Centre for High Performance Computing, South Africa, under the project ASTR0945.

\bibliographystyle{JHEP}
\bibliography{ref}

\end{document}